\documentstyle[12pt,psfig]{article}
\let\section=\subsection     \let\subsection=\subsubsection                %%
\newcommand{\doe}
{This work was supported by the
Director, Office of Energy Research,
Office of High Energy
and Nuclear Physics,
Division of Nuclear Physics,
of the U.S. Department of Energy under Contract
DE-AC03-76SF00098.}
\newcommand{\lbl}{LBNL-41322}

\newcommand{\eos}{equation of state~}

\newcommand{\eosp}{equation of state}
\newcommand{\Eos}{Equation of state}

\newcommand{\GR}{General Relativity~}

\newcommand{\GRp}{General Relativity}

\newcommand{\beqn}{\begin{eqnarray}}
\newcommand{\eeqn}{\end{eqnarray}}
\newcommand{\br} {\mbox{\mbox{\boldmath$r$}}}

\newcommand{\tit}
{PHASE TRANSITION SIGNAL IN PULSAR TIMING}
\newcommand{\autha} {Norman K. Glendenning}
\newcommand{\dateofdoc}{\today}
\newcommand{\adra}
{Nuclear Science Division \&
Institute for Nuclear and Particle Astrophysics,\\
  Lawrence Berkeley  National Laboratory,\\
   MS: 70A-3307, Berkeley, California 94720}
                                    
\begin{document}

%%\input title

%%%%%%%%%%%%%%%%%  TITLE PAGE %%%%%%%%%%%%%%%%%%%%%%%%%
\begin{titlepage}
%%\lbl
\parbox{2.5in}{\begin{flushleft}Hirschegg, Austria\end{flushleft}}%
\hfil
\parbox{2.5in}{\begin{flushright} \lbl \end{flushright}}
~\\[2ex]
\begin{center}
\begin{Large}
\renewcommand{\thefootnote}{\fnsymbol{footnote}}
\setcounter{footnote}{1}
\tit {\footnote{\doe}}\\[5ex]
\end{Large}

\renewcommand{\thefootnote}{\fnsymbol{footnote}}
\setcounter{footnote}{2}
\begin{large}
\autha \\[2ex]
\end{large}
\adra \\[2ex]
\dateofdoc \\[2ex]
\end{center}

%%%%%%%%%%%%%%%%% FOR NOTATION OF WHERE PRESENTED USE THE FOLLOWING
\begin{quote}
\begin{center}
{\bf Invited Paper, Hirschegg '98,  Nuclear Astrophysics \\
Organizers: M.\ Buballa, W.\ Norenberg, 
J.\ Wambach and A.\ Wirzba}
\end{center}
\end{quote}

%%%%%%%%%% front cover picture ..........................
%%%%%%%%%%%%%%%%%%%%%%%%%%%%%%%%%%%%%%%%%%%%%%%%%%%%%%%%%%%%%%%%%%%%%%%%%

\begin{figure}[tbh]
\vspace{-.5in}
\begin{center}
\leavevmode
\centerline{ \hbox{
\psfig{figure=ps.h4,width=2.5in,height=3in}
\hspace{.5in}
\psfig{figure=ps.h6,width=2.5in,height=3in}
}}
\end{center}
\end{figure}
%%%%%%%%%%%%%%%%%%%%%%%%%%%%%%%%%%%%%%%%%%%%%%%%%%%%%%%%%%%%%%%%%%%%%%%%%%
%%\begin{center}
%%{\bf PACS} 26.60+c,~97.10.Cv,~97.60.Jd,~12.39.Ba \\[4ex]
%%\end{center}
\end{titlepage}

%%%%%%%%%%%%%%%%%%%%First Page of Doc%%%%%%%%%%%%%%%%%%%%%%%%%%55
\begin{center}
   {\large \bf \tit}\\[5mm]
   Norman K. Glendenning \\[5mm]
   {\small \it  \adra \\[8mm] }
\end{center}

\begin{abstract}\noindent
A phase transition in the nature of matter in the core of a
neutron star, such as quark deconfinement or Bose condensation, can
cause the spontaneous spin-up of a solitary millisecond pulsar.
The spin-up epoch for our model lasts for $2\times 10^7$ years or 1/50
of the spin-down time (Glendenning, Pei and Weber in Ref. \cite{glen97:a}).
The possibility exists also for future measurements on X-ray neutron stars
with low-mass companions for mapping out the tell-tale ``backbending''
behavior of the moment of inertia. Properties of phase transitions in 
substances such as neutron star matter, which have more than one conserved
charge, are reviewed.
\end{abstract}

\section{Introduction}

Neutron stars have a high enough interior density as to make phase
transitions in the nature of nuclear matter a distinct possibility.
Examples are hyperonization, negative Bose condensation (like $\pi^-$
and $K^-$) and quark deconfinement.
According to the QCD property of asymptotic freedom, the most
plausible is the quark deconfinement transition. From   lattice
QCD simulations, this phase transition is expected to occur in very
hot ($T\sim 200$ MeV) or cold but dense matter. 
In this work we will use the deconfinement transition as an example,
but in principle, any transition that is accompanied by a sufficient
softening of the \eos and occurs at or near the limiting mass star,
can produce a similar signal. 

The paper is organized as follows. We discuss first the physical 
reason why a rapidly rotating pulsar, as it slows down over millions of
years  because of 
angular momentum loss through the weak electromagnetic process of 
magnetic dipole radiation, will change in density due to weakening centrifugal 
forces  and possibly
encounter, first at its center, and then in a slowly expanding region,
the conditions for a phase transition. 
Conversely, an accreting star will be spun up from low to high
frequency by accretion from a low-mass companion. This too will have a very 
long time-scale because accretion is regulated by the radiation
pressure of the star's surface, heated by infalling matter. 

After having discussed the reasons why we might see signals of
phase changes, both in rapidly rotating stars that are spinning down 
because of angular momentum loss to radiation and stars that are spinning up
due to the input of angular momentum by accretion, we discuss some
aspects of phase transitions that are common to all first order
transitions in neutron star matter, or more generally in isospin
asymmetric matter.

\section{Effects of Phase Transitions on Rotating Stars}

\subsection{Evolutionary Path of Neutron Stars}

Since neutron stars are born with  almost the highest density that
they will have in their lifetime, being very little deformed by
centrifugal forces, they will possess cores of the high density phase
essentially from birth if the critical density falls in the range of
neutron stars. However the global properties, such as mass or size,  of
a slowly rotating neutron star are little effected by whether or not
it has a more compressible phase in the core. In principle, cooling
rates should depend on interior composition, but cooling calculations
are beset by many uncertainties and competing assumptions about
composition can yield similar  cooling rates depending on other
assumptions about superconductivity and the cooling processes. 
Moreover,  for those stars
for which a rate has been measured, not a single mass is known. It is
unlikely that these measurements will yield conclusive evidence in the
present state of uncertainty \cite{page97:a,weber97:a}.

\begin{figure}[htb]
%%%%\vspace{-.4in}
\begin{center}
\leavevmode
\psfig{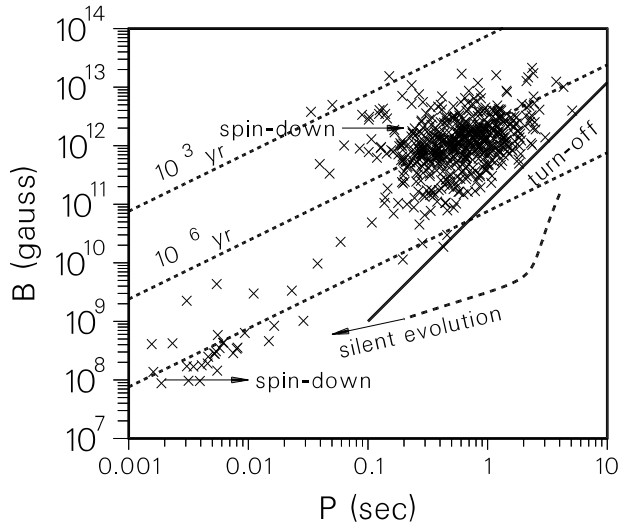}
\parbox[t]{4.6 in} { \caption { \label{pulsar} The 
evolutionary track  of
pulsars are from high magnetic field and moderate rotation period
to long period in about $10^7$ to $10^8$ years, to accreting
X-ray neutron stars,  to millisecond pulsars with low magnetic fields.
}}
\end{center}
%%%%%%\vspace{.2in}
\end{figure}
Nevertheless, it may be possible to observe the phase transition in
millisecond pulsars by the easiest of measurements---the sign of
$\dot{\Omega}$. Normally the sign should be negative corresponding to loss of
angular momentum by radiation. However a phase
transition that occurs near or at the
limiting mass star, can cause spin-up during a substantial era
compared to the spin-down time of millisecond pulsars.
The transition may be of either first or second order provided that
it is to an appreciable more compressible phase.  We sketch the
conventional evolutionary history \cite{heuvel91:a} of pulsars with the
addition of the supposition that the critical density for quark
deconfinement falls in the density range spanned by neutron stars.

As already remarked, with the supposition above, the star has a quark core
from birth but its properties are so little effected that this fact
cannot be discerned in members of  the canonical pulsar population. 
It is born with moderate rotation period, acquired by the conservation of
angular momentum during core collapse and with high magnetic field by
flux conservation. In a diagram of magnetic field strength $B$ and rotation
period $P$  (Fig.\ \ref{pulsar}),
it is injected near the line marked $10^3$ years (since birth) 
and with $B\sim 10^{13}$ gauss.  It evolves quickly at
constant $B$ toward longer period
for $\sim 10^6$  to $10^8$ years.
There it   lingers with the bulk of the pulsar population
at long periods (because $\dot{P}\sim 1/P$)
 before the combination of field strength and 
rotation
period are insufficient to accelerate  charged
particles  that 
produce the radiation. At that time the
star has entered the  radio silent epoch. 
Some pulsars  will have
had a less dense companion or will acquire one from which they
accrete matter and angular momentum. Some will be seen as X-ray emitters
during the radio silent  phase. During spin-up to frequencies much higher
than those with which pulsars are born,
the neutron star becomes increasingly centrifugally 
deformed and its interior density
falls. Consequently,  the radius at which the critical phase
transition  density occurs moves toward the center of the
star---quarks that were deconfined at the birth of the star,
recombine to form hadrons.
(This era may also be detectable as discussed
in section \ref{xray}.) When accretion ceases, and if
the neutron star has been spun up to a state in which the combination
of reduced field strength (perhaps to ohmic decay)
 and increased frequency turn the dipole
radiation on again, the pulsar recommences spin-down as a radio visible
millisecond pulsar.

During spin-down as   a millisecond pulsar,
 the central density increases with decreasing
centrifugal force. First at the center of the star, and then in an
expanding region, the highly compressible quark matter will replace
the less compressible nuclear matter. The quark core, weighed down by
the overlaying layers of nuclear matter is compressed to high density,
and the increased central concentration of mass acts on the overlaying
nuclear matter, compressing it further
(see Figs.\ \ref{prof_k300b180_log} and \ref{omega_r_k300b180_d}).
 The resulting decrease in the
moment of inertia  causes the star to spin up
to conserve angular momentum not carried off by radiation. The
phenomenon  is analogous to that of ``backbending'' predicted for
rotating nuclei by Mottelson and Valatin  \cite{mottelson60:a}
 and discovered in the 1970's 
\cite{johnson72:a,stephens72:a}
(see Fig.\ \ref{nucleus}). In
nuclei, it was established that the change in phase is from a particle
spin-aligned state at high  nuclear angular momentum to a superfluid
state at low angular momentum.  The phenomenon   is also analogous to
an ice skater who commences a spin with arms outstretched. Initially
spin decreases because of friction and air resistance, but a period of
spin-up is achieved by pulling the arms in. Friction then
reestablishes spin-down. In all three examples, spin up is a
consequence of a decrease in moment of inertia beyond what would occur
in response to decreasing angular velocity.
%%%%%%%%%%%%%8888888888888888888888888888888888888888888
\begin{figure}[tbh]
\vspace{-.5in}
\begin{center}
\leavevmode
\centerline{ \hbox{
\psfig{figure=ps.h2,width=2.7in,height=3.24in}
\hspace{.1in}
\psfig{figure=ps.h3,width=2.7in,height=3.24in}
}}
\begin{flushright}
\parbox[t]{2.5in} { \caption {Energy density profiles star  at
three rotation rates. Notice the 5 km radius quark matter core
when the star is not rotating (or only slowly as a canonical pulsar).
Mixed phase extends to 8 km.
\label{prof_k300b180_log}
}} \ \hspace{.3in} \
\parbox[t]{2.5in} {  \caption{Radial boundaries between different phases
for stars of mass indicated on the y axis.
Composition consists of quarks, the baryon octet
and leptons. The geometric phases
will be discussed in  Section \protect\ref{tran}.
\label{omega_r_k300b180_d}
}}
\end{flushright}
\end{center}
\end{figure}
%%%%%%%%%%%%%8888888888888888888888888888888888888888888

\subsection{Calculation}
In our calculation, nuclear matter
was described in a relativistically covariant theory
\cite{garpman79:a,glen85:b,glen91:c}
 and quark matter
in the MIT bag model
\cite{chodos74:a}. The phase transition occurs in a substance of
two conserved quantities, electric charge and baryon number, and must
be found in the way described in Ref. \cite{glen91:phase} and in 
Section \ref{tran}.
The moment of inertia
must incorporate all effects described above---changes in composition
of matter,  centrifugal stretching---and frame dragging,
all within the framework of \GRp. The expression derived by
Hartle is inadequate because it neglects these effects
\cite{hartle67:a,hartle67:b}. Rather
we must use the expression derived by us 
\cite{glen92:b,glen93:a}.

\begin{figure}[tbh]
\vspace{-.5in}
\begin{center}
\leavevmode
\centerline{ \hbox{
\psfig{figure=ps.h4,width=2.5in,height=3in}
\hspace{.5in}
\psfig{figure=ps.h5,width=2.5in,height=3in}
}}
\begin{flushright}
\parbox[t]{2.5in} { \caption {Nuclear moment of inertia as a function of squared
frequency for $^{158}$Er, showing backbending in the nuclear case.
\label{nucleus}
}} \ \hspace{.3in} \
\parbox[t]{2.5in} {  \caption{\Eos~for the fist order
deconfinement phase transition described in the text.
\label{eos_k300_y_h}
}}
\end{flushright}
\end{center}
\end{figure}

%%%%%%%%%%%%%%%%%%%%%%%%%%%%%%%%%%%%%%%%%%%%%

For   fixed baryon number we solve \GR for a star rotating at
a sequence of angular velocities corresponding to an \eos that describes
the deconfinement phase transition from  charge neutral nuclear matter
to quark matter. The \eos is shown in Fig.\ \ref{eos_k300_y_h}.
 The moment of inertia as a function of angular velocity
does not decrease monotonically as it would for a gravitating fluid of
constant composition. Rather, as described above, the epoch over which
an enlarging central region of the star enters the more compressible
phase is marked by spin-up (Fig.\ \ref{oi}).
Does the spin-up epoch endure long enough to provide a reasonable chance
of observing it some members of the pulsar population? This question
we turn to now.

%%%%%%%%%%%%%%%%%%%%%%%%%%%%%%%%%%%%%%%%%%%%%%%%%%%%%%%%%%

\subsection{Spin-up Era}
To estimate the duration of the spin-up, we solve the deceleration
equation for the star with moment of inertia having the behavior shown
in
Fig.\ \ref{oi}. 
From the energy loss equation 
 \begin{eqnarray}
 \frac{dE}{dt} =
  \frac{d}{dt}\Bigl(\frac{1}{2} I \Omega^2 \Bigr) =
   - C \Omega^{4}
    \label{energyloss}
      \end{eqnarray}
        for  magnetic dipole radiation we find
\begin{eqnarray}
\dot{\Omega}= -\frac{C}{I(\Omega)}
  \biggl[1  + \frac{I^{\prime}(\Omega) \,
\Omega}{2I(\Omega)}
 \biggr]^{-1} \Omega^3\,.
   \label{braking2}
\end{eqnarray}
This expression reduces to the usual braking equation when the moment
of inertia is held fixed. The braking index is a dimensionless combination
of three quantities that are observable in principle, namely the
angular velocity and its first two time derivatives. It
is generally thought to have a constant value, namely 3, 
for magnetic dipole radiation. However constancy would follow only if 
the star rotated rigidly.
Instead it  varies
with angular velocity and therefore time according to
  \begin{eqnarray}
  n(\Omega)\equiv\frac{\Omega \ddot{\Omega} }{\dot{\Omega}^2}
  = 3   - \frac{ 3  I^\prime \Omega +I^{\prime \prime} \Omega^2 }
    {2I + I^\prime \Omega}
     \label{index}
      \end{eqnarray}
      where $I^\prime \equiv dI/d\Omega$ and $I^{\prime\prime}
      \equiv dI^2/d\Omega^2$. This holds in general even for a star whose internal composition does not change with angular velocity (inconceivable). In particular one can see that for very high frequency, the derivatives will be largest and the braking index for any millisecond pulsar near the Kepler frequency will be less than the dipole value of $n=3$.

\begin{figure}[tbh]
\vspace{-.5in}
\begin{center}
\leavevmode
\centerline{ \hbox{
\psfig{figure=ps.h6,width=2.5in,height=3in}
\hspace{.5in}
\psfig{figure=ps.h7,width=2.5in,height=3in}
}}
\begin{flushright}
\parbox[t]{2.5in} { \caption { Moment of inertia of corresponding
to a change of phase.
Time flows from large to small
$I$.
\label{oi} 
}} \ \hspace{.3in} \
\parbox[t]{2.5in} { \caption {The time evolution of
the braking index plotted over two
decades that include the epoch of the phase transition.
\label{nt}
}}
\end{flushright}
\end{center}
\end{figure}
The braking index is shown as a function of time in Fig.\ \ref{nt}.
An anomalous value endures for $10^8$ years corresponding
to the slow spin down of the pulsar and the corresponding slow
envelopment of a growing central region by the new phase. The two
points that go to infinity correspond to the infinite derivatives of
$I$ at which according to (\ref{braking2}), the deceleration 
vanishes.
They mark the boundaries of the spin-up era.
The actual spin-up lasts for $2\times 10^7$ years or 1/50 of the
spin-down time for this pulsar. This could be easily observed in a
solitary pulsar and would likely signal a phase transition.

\subsection{X-Ray Pulsars in Low Mass Binaries}\label{xray}

As Lamb has discussed in this volume, neutron stars that 
are accreting mass that is channeled to their surface
from  a low-mass companion have been observed to rotate 
in the millisecond range.  
One of the goals in those studies is to observe phenomena associated with the last stable orbit. 
Mass, radius, pulsar frequency and the frame dragging frequency 
can  be determined in principle, and it is the goal of the experiments to do so.
From the combination of the frame dragging frequency and radius, 
the moment of inertia
can be obtained as
\beqn
I=\frac{1}{2} \frac{\omega}{\Omega}R^3~~~~~({\rm gravitational~units})
\eeqn
where the frame dragging angular velocity $\omega(R)$ corresponds to 
$R$, and $\Omega$ is the rotational angular velocity of the pulsar.
A few such sets of data  corresponding to stars of the same mass
would represent points
on a plot of moment of inertia vs. angular velocity, as in Fig.\ \ref{oi}. 
If indeed quark matter cores exist in slower pulsars, 
then the possibility exists that  some of the X-ray emitters will lie 
on the upper branch and others on the lower branch. Even if one did not 
have  observations that
lay on the backbend, observations that established the existence
of two 
assymptotes would be quite convincing evidence of a different 
phase on the two branches.

\section{Properties of Phase Transitions}\label{tran}

We discuss briefly phase transitions of interest in nuclear and nuclear astro physics. These include pion condensation \cite{migdal78:a}, 
hyperonization \cite{glen85:a}, 
kaon condensation \cite{kaon}, quark deconfinement in stars 
\cite{quark0} and recently, H-dibaryon Bose condensation \cite{schaffner}. 

\subsection{Maxwell Construction}
Depending on the model or strength of coupling, the above  transitions
 may be first or second order. Until recently \cite{glen91:a} 
first order phase transitions in nuclear matter or neutron star matter (nuclear matter in equilibrium under the constraint of charge neutrality), where implemented with some variant of the Maxwell construction (eg.  tangent slope ($\mu=d\epsilon/d\rho$)). Obviously the Maxwell construction can make only one 
chemical potential common to phases in equilibrium, and is a valid
construction for simple substances with one independent component,
such as water. However,  nuclear 
systems have two or
more
 conserved charges (baryon and electric charge in the case of 
neutron star matter or low density nuclear matter, and
strangeness in addition on the time-scale of high-energy reactions). 
For such substances, a
 little reflection  reveals that the chemical potential that is rendered 
common in the Maxwell construction is neither the baryon nor charge chemical 
potential but a varying combination according to the varying composition of 
matter as a function of baryon density.

The Maxwell construction causes the pressure to be  constant  in the mixed phase.
(This is obvious from the tangent construction in which the energy is a linear function of volume, $E=-pV+\mu N$ (with $p$ and $\mu$ constants of the construction) and hence the pressure, $-dE/dV$, is constant.) The consequence for stellar structure is striking: the mixed phase is absent in the monotonic pressure environment of 
a star and there is a large density discontinuity at the radial point 
corresponding to the constant pressure of the mixed phase.
Inside this radial point the dense quark matter resides; outside is the less dense nuclear matter.

In some work, phase transitions were implemented in a different but equivalent fashion. Charge neutrality was enforced by requiring the charge density to vanish identically. This is valid
 in either pure uniform phase. However when applied to the mixed phase 
it is too stringent a way of enforcing charge neutrality. All that is 
required by the balance of Coulomb and gravitational forces is that the star 
be charge neutral to a high degree ($Z_{{\rm net}}/A \leq (m/e)^2 
\approx 10^{-36}$); not that the
charge
 density vanishes identically.  {\sl Global} neutrality ($\int q(r) dV =0$) is all that is required. Indeed, when the details are examined, it will be seen that {\sl local} neutrality is incompatible with Gibbs criteria for phase equilibrium. Gibbs conditions and the conservation laws
can be satisfied
simultaneously 
only when the conservation laws are imposed
in a global sense.

\subsection{Gibbs Equilibrium}
We briefly review how to find the conditions for phase equilibrium in substances of more than one conserved charge that are in accord with Gibbs criteria for chemical, mechanical and thermal equilibrium \cite{glen91:a}. 
For definiteness
we consider a system with two conserved charges (or independent
components), namely, baryon number and electric charge number and refer to the phases as  1 and 2. Gibbs conditions
are summarized in
\beqn
p_1(\mu_n,\mu_e)=p_2(\mu_n,\mu_e)
\label{pres}
\eeqn
where $\mu_n$ and $\mu_e$ are the chemical potentials corresponding to
baryon number and electric charge. We understand that temperature $T$ is
held fixed. (It is 
small on the nuclear scale within several seconds of birth of a 
neutron star, and can set it
to zero.)
The above equation
must hold in conjunction with expressions for the conservation
of baryon and charge number, $B$ and $Q$. 
The unknowns are the two chemical potentials and the volume $V$ of the 
sample containing the charges.
($V$ is not the volume of the star but any {\sl locally inertial}
volume in the star, one in which the laws of special relativity 
hold to high precision \cite{book}.) For a volume 
fraction $\chi=V_2/V$ of phase 2, the conditions of
global conservation
can be expressed (for a uniform region)
as
%%%%%%%%%%%%%%%%%%%%%%%% insert equations
  \beqn
  & & \frac{1}{V} \int_V \rho(\br) d\br =
   (1-\chi) \rho_1(\mu_n,\mu_e)+ \chi \rho_2(\mu_n,\mu_e)\equiv
\frac{B_1+B_2}{V} =
   \frac{B}{V}~~ \label{dens} \\
   & & \frac{1}{V} \int_V q(\br) d\br =
 (1-\chi) q_1(\mu_n,\mu_e)+ \chi q_2(\mu_n,\mu_e) \equiv
\frac{Q_1+Q_2}{V} =\frac{Q}{V}~~
   \label {neut}
   \eeqn
   where $B_{1,2}$ and $Q_{1,2}$ are the baryon and charge numbers, 
$\rho_{1,2}$ and $q_{1,2}$ 
   are the baryon and charge densities in the volumes 
$V_1$ and $V_2$ occupied respectively by the two phases.
     The above three equations  (\ref{pres}-\ref{neut})
     serve to
     determine the two independent chemical potentials
     $\mu_n ,~\mu_e$ 
     and volume
    $V$
      for a {\sl specified}
      volume fraction $\chi$ of phase `2' in equilibrium with phase `1'.
      Thus the solutions are of the form
      \beqn
      \mu_n=\mu_n(\chi),~~~~\mu_e=\mu_e(\chi),~~~~V=V(\chi)\,.
      \eeqn
The equilibrium condition (\ref{pres}) therefore can be rewritten as
\beqn
p_1(\chi)=p_2(\chi)\,.
\eeqn
      This shows, as concerns the bulk 
properties, that the common pressure and all properties of the
      phases in equilibrium vary as the proportion $\chi$ and that the
      pressure of a multi-component system in the mixed phase is not
in general constant. These are fundamentally different properties for
phase equilibrium of multi-component substances; they contrast with the
properties of single-component substances such as water, in which the 
properties are independent of the proportion of the phases.
For nuclear systems, the only exception to the above conclusion is for
symmetric nuclear matter. In that case the system, by preparation, is optimum,
and no rearrangement of conserved charges will take place.

\subsection{Internal Driving Forces}

We discuss now the microphysics responsible for variation of all properties
of the phases in equilibrium as their proportion varies.
      It is clear from the above discussion that there is a degree 
(or degrees)  of
freedom in a multi-component substance that can be exploited by the internal forces  to lower the energy.  To see this, consider the concentration of the conserved quantities in both of the pure phases.
It is some definite
      number
      \beqn
	    c={Q}/{B}
		  \eeqn
      according to the way the system was prepared whether 
      in a test tube by a chemist,
      or in a neutron star by nature through the partially chaotic
      processes of a supernova.
      The degree(s) of freedom that the system can exploit
      to find the energy minimum
      in the mixed phase is that of rearranging the
      concentration of the conserved charges
      in each phase in equilibrium
      \beqn
      c_1=  {Q_1}/{B_1}\,,~~~~~
      c_2=  {Q_2}/{B_2}
      \eeqn
      subject to the overall conservation laws (\ref{dens},\ref{neut}).
      More generally, if the system is composed of $n$ conserved charges,
      there are $n-1$ such degrees of freedom.
 In particular,
      a single component 
      substance does not possess any freedom which is why the pressure
      and all properties of the two phases remain the same for all proportions
      of the phases in equilibrium (like water and ice).

In nuclear 
matter
 the internal force that drives the redistribution of charge 
allowed by the conservation laws is the isospin symmetry force that is 
responsible for the valley of beta stability. About half the symmetry
energy arises 
from the Fermi energy and (in our model) the other half to the 
coupling of isospin to the $\rho$ meson.
Neutron star matter is highly isospin asymmetric. When conditions (say of increasing pressure toward the center of the star)  cause a small amount of nuclear matter to transform to the other phase, the isospin driving force will exchange charge between regions of the two phases so as to make the neutron star matter more symmetric (positively charged) to the extent permitted by the conservation laws. The other phase will have a corresponding negative charge. The 
scope for exchanging charge, changes with the
fraction of new phase
(hence the non-linearity of energy with volume and the variation
of pressure with volume). We show in Fig.\ \ref{chiq_k300b180}
how the charge density in each phase varies as their proportion while the total charge is zero. (In the special case of isospin symmetric matter, the concentration is already optimum and the pressure will not vary in the mixed phase.) 
%*****************************************************

\begin{figure}[tbh]
\vspace{-.4in}
\begin{center}
\leavevmode
\centerline{ \hbox{
\psfig{figure=ps.h8,width=2.5in,height=3in}
\hspace{.5in}
\psfig{figure=ps.h9,width=2.5in,height=3in}
}}
\begin{flushright}
\parbox[t]{2.5in} { \caption { \label{chiq_k300b180}The charge
density carried by regions of confined and deconfined phases, and on leptons, 
assumed to be uniformly distributed. Densities
times the respective volume fraction add to zero.
}} \ \hspace{.3in} \
\parbox[t]{2.5in} { \caption { \label{size} Size (S $=2r$)
 and spacing (D $=2R$)  of geometrical
structures. Notation `q rods' means quark rods of negative charge
immersed in nuclear matter of positive charge. Charge densities of each 
phase as a function of proportion $\chi$ of quark phase are shown in Fig.\ 
\protect\ref{chiq_k300b180}.
}}
\end{flushright}
\end{center}
\end{figure}

The case in which electric charge is one of the conserved quantities is special.   Because Coulomb is a long-range force, charges will arrange themselves so as not to create large volumes of like charge. Regions of the two oppositely
charged phases will tend to shield each other.
The surface interface energy resists breakup into small regions. The competition will define the dimensions of charged regions and their spacing as described next.

\subsection{Spatial Structure}

To calculate the spatial order we
use the Wigner-Seitz {\sl approximation} by choosing a volume $v$ which is the cell size
and contains the rare phase of dimension $r$ and the dominant
phase in such amount as makes the cell neutral. Therefore cells do 
not interact.
The Coulomb and surface energies per unit volume
can be written in the schematic form that shows their dependence 
on dimension $r$ of the ``geometry'' and on the proportion $\chi$.
\beqn
E_C/v=C(\chi)r^2,~~~~~E_S/v=S(\chi)/r\,.
\label{cs}
\eeqn
(The dependence on $r$ can be obtained by dimensional analysis
\cite[See end of Section V]{glen91:a}.
Minimizing  their sum as a function of $r$ at fixed $\chi$ yields
\beqn
E_S=2E_C
\eeqn
as always happens in the minimization of a 
sum of two quantities that vary as $r^2$ and $1/r$ respectively.
In the above, $C$ and $S$ are specific functions of proportion $\chi$ whose
form is dictated by the geometry of the cells (eg. sphere, rods and slabs) but for brevity we do not write them down.
The above equations serve to define the droplet radius  and cell size for each proportion of the phases, 
\beqn
r= \Bigl(\frac{S(\chi)}{2C(\chi)}\Bigr)^{1/3},~~~~~R=\frac{r}{\chi^{1/3}}\,,
\eeqn
where for spherical geometry, $\chi=(r/R)^3$.
As remarked in the introduction, the internal force that drives the
charge redistribution between phases in equilibrium is the 
isospin restoring force. As one can see, the size and spacing of the
droplets of rare phase immersed in the dominant will vary as proportion
$\chi$. Other geometries besides spheres may minimize the energy
according to the proportion.
The functional form of $S$ and $C$ is distinct in each case, as is the relation of the dimensions of $r$ and $R$ to $\chi$. 
\beqn
& & C_d(\chi) =2\pi \bigl\{[q_1(\chi)-q_2(\chi)] e\bigr\}^2  \chi f_d(x)\\
& & S_d(\chi) =\chi \sigma d
\eeqn
where $d=1,2,3$ for the idealized geometries of slabs, rods and drops
respectively, $\sigma$ is the surface tension and
\beqn
x\equiv (r/R)^d\,,~~~~~d=1,2,3\,
\label{chir}
\eeqn
where $x$ is related to the proportion $\chi$ by
\beqn
x= \left\{ \begin{array}{ll}
 \chi  & \mbox{,background~phase~is~1} \\
1-\chi & \mbox{,background~phase~is~2}
\end{array} \right.
\eeqn
and 
\beqn
f_d(x) =\frac{1}{d+2}
\biggl[\frac{1}{(d-2)}(2-d x^{1-2/d}) + x \biggr]\,.
\eeqn
The above considerations 
 are identical to those encountered in a description of nuclei embedded in an electron gas. At higher relative concentration of nuclear matter to electron gas, the spheres will merge to form rods and so on \cite{ravenhall83:aa}. 
What is remarkable in the present context is that two phases of one and the same substance, under the action of the isospin symmetry 
restoring force, are endowed with opposite charge and form a Coulomb lattice.
The size and spacing and the geometric form that minimizes the sum of
surface and Coulomb energies are shown in Fig.\ \ref{size}.

\subsection{Three Theorems}

We have thus three theorems concerning the equilibrium configuration
of the mixed phase
of  a first order phase transition
of a substance 
with more than one conserved charge (or independent component in the 
language of chemistry):
\begin{enumerate}
\item
All properties of the phases in equilibrium, including common pressure
vary as the proportion of phases.
\item
If electric charge is one of the conserved charges, the mixed phase
will be in the form of a crystalline lattice.
\item
Because of theorem 1, the geometry of the crystal
and the size and the spacing of the
lattice will vary with proportion.
\end{enumerate}

These remarkable properties of first order
phase transitions 
and the role played by the microphysics or internal forces
is discussed in detail elsewhere
\cite{glen91:a,book,glen94:e}.
It will be observed that the above discussion is completely general,
and must apply to many systems in physical chemistry, nuclear physics,
astrophysics and cosmology. In
particular, in nuclear systems it applies 
to the confined-deconfined
phase transition at high density, to the so-called
liquid-vapor transition  at sub-saturation density as well as to Pion
and Kaon condensation if they are first order transitions, and if they
occur. (It is understood that if a phase transition is induced by a nuclear collision, complete equilibrium will not be achieved, and in particular,  there is insufficient time for formation of spatial structure.)

One step in the calculation remains to be described. The sum of surface and Coulomb energies, requires a knowledge of certain bulk properties like the charge density of each phase in equilibrium. The bulk energy and pressure can be computed according to ones favorite theory, and phase equilibrium finally has to be computed self-consistently, by minimizing the sum of all three energies, which can conveniently be done by iteration. The surface tension is generally not known and needs to be
computed self-consistently for the phases in equilibrium. It will be a function of proportion, just as all other properties are.

Of course we have properly referred to the calculation as approximate:
the division of total energy into bulk, surface and Coulomb is approximate. If one has reason to question the range of densities spanned by the structured mixed phase, then the alternative is to compute the energy on a lattice.

There is another important consequence
of the existence 
of  degree(s) of freedom for rearranging
concentrations of conserved quantities  in
multi-component substances according to the energy
minimization principle. Whereas in one-component substances the properties
of each phase in equilibrium are very unlike (as for example the 
density of ice and water), in a multi-component substances the
rearrangement of charges so as to optimize the energy at each proportion
of the phases relaxes their differences. This can be seen
in Fig.\ 7 of Ref.\ \cite{glen91:a}. 
As a consequence,
the transition density from the pure low-density phase to the 
mixed phase
is lower than would be expected were the degree(s) of freedom frozen
out (as in the pre-1990 studies of deconfinement in neutron stars)
\cite{glen91:a,glen95:c,heiselberg93:a,pandharipande94:a}.

\subsection{Summary}

A change in phase  during the course of spin-down of a millisecond pulsar occasioned by rising internal density due to weakening centrifugal forces will be reflected in a change in moment of inertia. The moment of inertia will follow different laws as a function of rotational frequency before and after the transition as in Fig.\ \ref{oi}. The transition from one law to the other 
can (and is in our example) pass through an era of spin-up. 
This era lasts for about 1/50 of the spin-down time of millisecond pulsars,
which represents an event rate. Spin up is trivial to detect and would be
spectacular in a solitary pulsar which ought to be spinning down because
of angular momentum loss to radiation.

For a weaker transition than 
experienced in our model  the spin-up region need not occur: 
the moment of inertia may simply change smoothly 
from one trajectory to the other.
In that case a 
sufficient number of observations of X-ray
emitters in low-mass binaries
could still identify a transition, even though 
the transition era were not actually observed, provided the 
observed neutron stars lay some on one branch, some on the other.

For distinct branches to exist, it appears  to be necessary that the phase 
transition occurs near the maximum mass. The 
transition can be first or second order as long as it is accompanied
by a sufficient softening of the \eosp.

 %....................BIBLIOGRAPHY.....................


\begin{thebibliography}{10}

\bibitem{glen97:a}
N. K. Glendenning, S. Pei and F. Weber, Phys.\ Rev.\ Lett.\ {\bf 79} (1997)
  1603.

\bibitem{page97:a}
D. Page, {\sl Thermal Evolution of Isolated Neutron Stars}, To be published in
  the proceedings of the NATO ASI `The Many Faces of Neutron Stars' (Kluwer),
  Eds A. Alpar, R. Buccheri \& J. van Paradijs.

\bibitem{weber97:a}
C. Schaab, B. Hermann, F. Weber and M. K. Weigel, Astrophys.\ J.\ Lett.\ {\bf
  480} (1997) L111.

\bibitem{heuvel91:a}
D. Bhattacharya and E. P. J. van den Heuvel, Physics Reports, {\bf 203} (1991)
  1 .

\bibitem{mottelson60:a}
B. R. Mottelson and J. G. Valatin, Phys.\ Rev.\ Lett.\ {\bf 5} (1960) 511.

\bibitem{johnson72:a}
A. Johnson, H. Ryde and S. A. Hjorth, Nucl.\ Phys.\ {\bf A179} (1972) 753.

\bibitem{stephens72:a}
F. S. Stephens and R. S. Simon, Nucl.\ Phys.\ {\bf A183} (1972) 257.

\bibitem{garpman79:a}
S. I. A. Garpman, N. K. Glendenning and Y. J. Karant, Nuc.\ Phys.\ {\bf A322}
  (1979) 382.

\bibitem{glen85:b}
N. K. Glendenning, Astrophys.\ J.\ {\bf 293} (1985) 470.

\bibitem{glen91:c}
N. K. Glendenning and S. A. Moszkowski, Phys.\ Rev.\ Lett.\ {\bf 67} (1991)
  2414.

\bibitem{chodos74:a}
A. Chodos, R. L. Jaffe, K. Johnson, C. B. Thorne and V. F. Weisskopf, Phys.\
  Rev.\ D {\bf 9} (1974) 3471.

\bibitem{glen91:phase}
N. K. Glendenning, Nuclear Physics B (Proc. Suppl.) {\bf 24B} (1991) 110; \\
  Phys. Rev. D, {\bf 46} (1992) 1274.

\bibitem{hartle67:a}
J. B. Hartle, Astrophys.\ J.\ {\bf 150} (1967) 1005.

\bibitem{hartle67:b}
J. B. Hartle and D. Sharp, Astrophys.\ J.\ {\bf 147} (1967) 317.

\bibitem{glen92:b}
N. K. Glendenning and F. Weber, Astrophys.\ J.\ {\bf 400} (1992) 647.

\bibitem{glen93:a}
N. K. Glendenning and F. Weber, Phys. Rev. D {\bf 50} (1994) 3836.

\bibitem{migdal78:a}
A. B. Migdal, Rev.\ Mod.\ Phys.\ {\bf 50} (1978) 107.

\bibitem{glen85:a}
N. K. Glendenning, Phys.\ Lett.\ {\bf 114B} (1982) 392; \\ N. K. Glendenning,
  Astrophys.\ J.\ {\bf 293} (1985) 470; \\ N. K. Glendenning, Z. Phys. A {\bf
  326} (1987) 57; \\ N. K. Glendenning, Z.\ Phys. A {\bf 327} (1987) 295.

\bibitem{kaon}
V. A. Ambartsumyan and G. S. Saakyan, Astron.\ Zh.\ {\bf 37} (1963) 193 [Soviet
  Ast.\ -- AJ,{\bf 4} (1960) 187];\\ Ya. B. Zel'dovich and I. D. Novikov, {\sl
  Relativistic Astrophysics, Vol. 1, Stars and Relativity} (University of
  Chicago Press, 1971); \\ N. K. Glendenning, Astrophys.\ J.\ {\bf 293} (1985)
  470; \\ D. B. Kaplan and A. Nelson, Phys.\ Lett.\ {\bf 175 B} (1986) 57; \\
  H. D. Politzer and M. B. Weise, Phys.\ Lett.\ B {\bf 273} (1991) 156; \\ G.
  E. Brown, H. Lee, M. Rho, and V. Thorsson, Nucl.\ Phys.\ A {\bf 567} (1994)
  937; \\ V. Thorsson, M. Prakash and J. M. Lattimer, Nucl.\ Phys.\ A {\bf 572}
  (1994) 693;\\ V. Koch, Phys.\ Lett.\ B {\bf 337} (1994) 7; \\ E. E.
  Kolomeitsev, D. N. Voskresensky and B. Kampfer, Nucl.\ Phys.\ A {\bf 588}
  (1995) 889; \\ N. Kaiser. P. B. Siegel and W. Weise, Nucl.\ Phys.\ A {\bf
  594} (1995) 325; \\ T. Wass, N. Kaiser and W. Weise, Phys.\ Lett.\ B {\bf
  379} (1996) 34;\\ J. Schaffner and I. N. Mishustin, Phys.\ Rev.\ C {\bf 53}
  (1996) 1416 .

\bibitem{quark0}
G. Baym and S. A. Chin, Phys.\ Lett.\ {\bf 62B} (1976) 241;\\ G. Chapline and
  M. Nauenberg, Nature {\bf 264} (1976) 235; Phys.\ Rev.\ D {\bf 16} (1977)
  456.\\ B. D. Keister and L. S. Kisslinger, Phys.\ Lett.\ {\bf 64B} (1976)
  117.

\bibitem{schaffner}
N. K. Glendenning and J. Schaffner, H-Dibaryon Bose Condensate in Compact Stars
  (in preparation) 1997.

\bibitem{glen91:a}
N. K. Glendenning, Phys. Rev. D, {\bf 46} (1992) 1274.

\bibitem{book}
N. K. Glendenning, {\sl COMPACT STARS, Nuclear Physics, Particle Physics, and
  General Relativity} (Springer--Verlag New York, 1997).

\bibitem{ravenhall83:aa}
D. G. Ravenhall, C. J. Pethick and J. R. Wilson, Phys.\ Rev.\ Lett.\ {\bf 50}
  (1983) 2066.

\bibitem{glen94:e}
N. K. Glendenning, {\sl A Crystalline Quark-Hadron Mixed Phase in Neutron
  Stars}, Physics Reports, {\bf 264} (1995) 143.

\bibitem{glen95:c}
N. K. Glendenning and S. Pei, Phys.\ Rev. C {\bf 52} (1995) 2250.

\bibitem{heiselberg93:a}
H. Heiselberg, C. J. Pethick, and E. F. Staubo, Phys.\ Rev.\ Lett.\ {\bf 70}
  (1993) 1355.

\bibitem{pandharipande94:a}
V. R. Pandharipande and E. F. Staubo, in {\sl Proc. 2'nd International Conf. of
  Physics and Astrophysics of Quark-Gluon Plasma,} Calcutta, 1993, Eds. B.
  Sinha, Y. P. Viyogi and S. Raha, (World Scientific, 1994).

\end{thebibliography}
\end{document}